\documentclass[a4paper,11pt]{article}
\usepackage{pos}
\usepackage{qcd2}

\title{TMD phenomenology motivated by nonperturbative structures}

\author*[a,b]{Ted Rogers}
\author[c]{F.~Aslan}
\author[d,e]{M.~Boglione}
\author[d,e]{J.~O.~Gonzalez-Hernandez}
\author[a]{T.~Rainaldi}
\author[a,b]{A.~Simonelli}

\affiliation[a]{Department of Physics, Old Dominion University, Norfolk, VA 23529, USA}

\affiliation[b]{Jefferson Lab, 12000 Jefferson Avenue, Newport News, VA 23606, USA}

\affiliation[c]{Center for Nuclear Femtography, SURA,
 12000 Jefferson Avenue, Newport News, VA 23606, USA}

\affiliation[d]{Dipartimento di Fisica, Universit\`a degli Studi di Torino, Via P. Giuria 1, I-10125, Torino, Italy}

\affiliation[e]{INFN, Sezione di Torino, Via P. Giuria 1, Torino, I-10125, Italy}

\emailAdd{trogers@odu.edu}

\abstract{This talk summarized work done recently to organize the steps for implementing TMD phenomenology in a way optimized for contexts where the extraction and interpretation of hadronic structures and nonperturbative effects is the primary driving motivation.}

\FullConference{7th International Workshop on “Transverse phenomena in hard processes and the transverse structure of the proton (Transversity2024)\\
 3-7 June 2024\\
Trieste, Italy\\}

\begin{document}
\maketitle

\section{Introduction}
\label{s.intro}

I will structure these proceedings as follows: To establish context, I will briefly review the basic setup of TMD factorization for Drell-Yan scattering. Next, I will review the most common steps used to construct parametrizations of TMD pdfs in phenomenological implementations and discuss some of their advantages and disadvantages. Then, I will summarize our proposed modifications from Refs.~\cite{Gonzalez-Hernandez:2022ifv,Gonzalez-Hernandez:2023iso,Aslan:2024nqg}. I will organize the remainder of the proceedings around answering questions that arose in the context of the talk. 

A familiar example of a TMD factorization formula is the one for unpolarized Drell-Yan scattering~\cite{Collins:2011qcdbook},
\begin{align}
& \frac{\diff{\sigma}}{\diff[4]{{q}{}} \diff{\Omega}} = \mathcal{H}(Q_0,Q_0/\mu_{Q_0}) \times \no
& \times \int \diff{^2 \T{k}{a}}{} \diff{^2 \T{k}{b}}{} f_{j/h_a}(x_a,\T{k}{a};\mu_{Q_0},Q_0^2) f_{\bar{\jmath}/h_b}(x_b,\T{k}{b};\mu_{Q_0},Q_0^2) \delta^{(2)} (\T{q}{} - \T{k}{a} - \T{k}{b}) \ , \label{e.basicTMDmodel}
\end{align}
with TMD pdfs $f_{j/h_a}(x_a,\T{k}{a};\mu_{Q_0},Q_0^2)$ and $f_{\bar{\jmath}/h_b}(x_b,\T{k}{b};\mu_{Q_0},Q_0^2)$ describing longitudinal momentum fractions and transverse momenta for the incoming partons in each hadron $h_a$ and $h_b$. There is also a hard factor $\mathcal{H}$ and a sum over all parton flavors. Apart from the dependence on auxiliary variables associated with evolution, \eref{basicTMDmodel} has exactly the form of a TMD parton model with a clear hadron structure interpretation.
(For brevity, power suppressed errors, sums over flavors, and a $Y$-term correction for $\Tsc{q}{} \approx Q_0$ are not written explicitly.) The renormalization and Collins-Soper (CS) scales are normally written as generic scales $\mu$ and $\sqrt{\zeta}$, but since we have physical applications in mind I express them with physical input scales $\mu_{Q_0}$ and $Q_0$, fixed to $\mu_{Q_0} \propto Q_0$ from the outset. 

It is often the case that one prefers to express \eref{basicTMDmodel} in terms of transverse coordinate space TMDs where implementing evolution becomes simple,
\begin{equation}
\label{e.basicTMDmodelbspace}
\frac{\diff{\sigma}}{\diff[4]{{q}{}} \diff{\Omega}} =  
\mathcal{H}(Q_0,Q_0/\mu_{Q_0}) \int 
 \frac{\diff[2]{\T{b}{}}}{(2 \pi)^2}
    ~ e^{i\T{q}{h}\cdot \T{b}{} }
    ~ \tilde{f}_{j/h_a}(x_a,\T{b}{};\mu_{Q_0},Q_0^2) 
    ~ \tilde{f}_{\bar{\jmath}/h_b}(x_b,\T{b}{};\mu_{Q_0},Q_0^2)  \, .
\end{equation}
The formulas above are written for the case of scales near the input scale, $Q_0$, which is understood to be the lowest scale for which TMD factorization is considered reasonable. To relate them to higher $Q$ requires solving evolution equations~\cite{Collins:2011qcdbook}, 
\begin{align}
 \frac{\partial  \ln \tilde{f}_{j/h}(x,{\Tsc{b}{}}{};\mu_Q,\zeta)}{\partial \ln \sqrt{\zeta} } ={}& \tilde{K}({\Tsc{b}{}}{};\mu)  \, , \label{e.CSeq}  \\
 \frac{\text{d}{ \tilde{K}({\Tsc{b}{}}{};\mu)}}{\text{d}{\, \ln \mu}{}} ={}&  - \gamma_K(\alpha_s(\mu))  \, , \label{e.CSrg} \\
\frac{\text{d}{\, \ln  \tilde{f}_{j/h}(x,{\Tsc{b}{}}{};\mu_Q,\zeta)}}{\text{d}{\, \ln \mu}} ={}&  \gamma(\alpha_s(\mu);\zeta /\mu^2) = \gamma(\alpha_s(\mu);1) - \gamma_K(\alpha_s(\mu)) \frac{1}{2} \ln \left(\frac{\zeta}{\mu^2}\right) \, . \label{e.TMDrg}
\end{align}
Here, $\gamma$ and $\gamma_K$ are perturbatively calculable anomalous dimensions and $\tilde{K}({\Tsc{b}{}}{};\mu)$ is the CS kernel, which is perturbatively calculable at small $\Tsc{b}{}$, but is nonperturbative in the large-$\Tsc{b}{}$ limit. In $\Tsc{b}{}$-space, the exact solution to \erefs{CSeq}{TMDrg} for all $\Tsc{b}{}$ is simple, and the factorized cross section at an arbitrary $Q$ becomes
\begin{align}
&\frac{\diff{\sigma}}{\diff[4]{{q}{}} \diff{\Omega}} =  
\mathcal{H}(Q,Q/\mu_{Q}) \no
& \int 
 \frac{\diff[2]{\T{b}{}}}{(2 \pi)^2}
    ~ e^{i\T{q}{h}\cdot \T{b}{} }
    ~ \tilde{f}_{j/h_a}(x_a,\T{b}{};\mu_{Q_0},Q_0^2) 
    ~ \tilde{f}_{\bar{\jmath}/h_b}(x_b,\T{b}{};\mu_{Q_0},Q_0^2)   E(\mu_Q/\mu_{Q_0},Q/Q_0,\mu_{Q_0} \Tsc{b}{})^2  \, . \label{e.basicTMDmodelbspace2}
\end{align}
with only a multiplicative evolution factor
\begin{align}
&E(\mu_Q/\mu_{Q_0},Q/Q_0,\mu_{Q_0} \Tsc{b}{})^2 \no
& \; = \exp\left\{  
        \tilde{K}(\Tsc{b}{};\mu_{Q_0}) \ln \parz{\frac{ Q^2 }{ Q_0^2}}
           +\int_{\mu_{Q_0}}^{\mu_Q}  \frac{ \diff{\mu'} }{ \mu' }
           \biggl[ 2 \gamma(\alpha_s(\mu'); 1) 
                 - \ln\left(\frac{Q^2}{ {\mu'}^2 } \right)\gamma_K(\alpha_s(\mu'))
           \biggr]
  \right\} \, .
\end{align}
(The square on $E$ is present because there is an evolution factor for each TMD pdf.) 
Therefore, knowledge of the TMD pdfs and the CS kernel at an input scale and for all $\Tsc{b}{}$, along with perturbative calculations of the anomalous dimensions, is enough to determine the TMD pdfs and the cross section at any higher scales. 

To obtain expressions for the TMD pdfs at $\Tsc{k}{} \approx Q_0$ or higher (or at $\Tsc{b}{} \lesssim 1/Q_0$ in coordinate space), one also normally takes advantage of the ability to use \emph{collinear} factorization to describe the perturbative tail,
or in terms of the small-$\Tsc{b}{}$ limit to use an operator product expansion, 
\begin{align}
& {\tilde{f}_{j/h}(x,{b_T}{};\mu_{Q_0},Q_0^2)} =  \int_x^1 \frac{\text{d} \xi}{\xi} \tilde{C}_{j/k}(x/\xi,b_T;\mu_{Q_0},Q_0^2,\alpha_s(\mu_{Q_0})) {f_{k/p}(\xi;\mu_{Q_0})} + O\left({b_T \Lambda_\text{QCD}}\right)^a \, ,  \label{e.ope}
\end{align}
where $\tilde{C}_{j/k}$ is a hard coefficient and $a > 0$. The only logarithmic behavior in the perturbative expression for $\tilde{C}_{j/k}$ is of the form $\ln^n \parz{\Tsc{b}{} Q_0}$, so using \eref{ope} directly in calculations is already optimal for $Q \approx Q_0$ where the only relevant perturbative region of $\Tsc{b}{}$ occurs at $\Tsc{b}{} \approx 1/Q_0$. By definition, all larger regions of $\Tsc{b}{}$ are to be considered nonperturbative. However, after one evolves to $Q \gg Q_0$, the $\Tsc{b}{} \ll 1/Q_0$ region starts to become important, and the $\ln^n \parz{\Tsc{b}{} Q_0}$ terms grow unacceptably large. To account for this, one normally performs another scale transformation on $\tilde{f}_{j/h}(x,{b_T}{};\mu_{Q_0},Q_0^2)$ in the $\Tsc{b}{} \ll 1/Q_0$ region from $(\mu_{Q_0},Q_0) \to \sim 1/\Tsc{b}{}$ via another application of \erefs{CSeq}{TMDrg}.  

The abstract set of steps summarized above are made concrete in applications by replacing the various pieces with a combination of perturbative approximations and nonperturbative parametrizations. The details of the different ways this is done in practice come with advantages and disadvantages depending on the specific goals, to be discussed below.   

\section{$\bstarsc$ method with $g$-function prescription}
\label{s.standard}

A basic task is to fix a method for merging nonperturbative descriptions of transverse momentum dependence with perturbative descriptions. In transverse coordinate space, that amounts to matching large-$\Tsc{b}{}$ nonperturbative descriptions of $\tilde{f}_{j/h}(x,{b_T}{};\mu_{Q_0},Q_0^2)$ and $\tilde{K}({\Tsc{b}{}}{};\mu)$ onto small-$\Tsc{b}{}$ perturbative calculations. 

The most common way of organizing the separation is the $\bstarsc$-prescription~\cite{Collins:1981va,Collins:1984kg}. It is motivated by a desire to separate out a factor that one can view as strictly perturbative, in the sense that it is only sensitive to small-$\Tsc{b}{}$ behavior and has zero or only minimal sensitivity to parameters that are meant to describe truly nonperturbative transverse momentum dependence.
In other words, the perturbatively describable transverse coordinate dependence is to be quarantined from explicitly nonperturbative models of transverse coordinate dependence. The way it is implemented is by introducing an arbitrary function that freezes $\T{b}{}$ when $\Tsc{b}{} \to \infty$,
\begin{equation}
\bstar(\Tsc{b}{}) = 
\begin{dcases}
\T{b}{} & b_T \ll b_{\rm max} \\
\vect{b}_{\rm max} & b_T \gg b_{\rm max} \,  \label{e.bdefold}
\end{dcases}\, ,
\end{equation}
with $\bmax$ understood to mark a boundary between what will be treated perturbatively and what will be considered nonperturbative regions of $\Tsc{b}{}$ dependence. (Thus, $\bmax \lesssim 1/Q_0$.)
The most common realization of \eref{bdefold} is 
\begin{align}
\label{e.bstar}
  \bstar(\Tsc{b}{}) = \frac{ \T{b}{} }{ \sqrt{ 1 + \Tscsq{b}{}/\bmax^2} } \, ,
\end{align}
though any function that satisfies \eref{bdefold} is acceptable, and others have been used in practice. 
The TMD pdfs at the input scale are then written as 
\begin{equation}
\tilde{f}_{j/h}(x,{\Tsc{b}{}}{};\mu_{Q_0},Q_0^2) = \tilde{f}_{j/h}(x,\bstarsc;\mu_{Q_0},Q_0^2) \parz{ \frac{\tilde{f}_{j/h}(x,{\Tsc{b}{}}{};\mu_{Q_0},Q_0^2)}{\tilde{f}_{j/h}(x,\bstarsc;\mu_{Q_0},Q_0^2)} }  \, . \label{e.firstratio}
\end{equation}
The factor in front is still just the coordinate space TMD pdf, but now it is only ever evaluated at $\Tsc{b}{} < \bmax$, so we anticipate using collinear factorization to calculate it. The factor in parentheses vanishes like a power of $\Tsc{b}{}$ at small $\Tsc{b}{}$, so we label it as being the ``nonperturbative'' contribution. It is conventionally written in the form of an exponential of a function, so that  
\begin{equation}
\tilde{f}_{j/h}(x,{\Tsc{b}{}}{};\mu_{Q_0},Q_0^2) = \tilde{f}_{j/h}(x,\bstarsc;\mu_{Q_0},Q_0^2) e^{- g_{j/h}(x,\Tsc{b}{}) }  \,  \label{e.expong}
\end{equation}
with
\begin{equation}
g_{j/h}(x,\Tsc{b}{}) \equiv - \ln \parz{\frac{\tilde{f}_{j/h}(x,{\Tsc{b}{}}{};\mu_{Q_0},Q_0^2)}{\tilde{f}_{j/h}(x,\bstarsc;\mu_{Q_0},Q_0^2)}} \, . \label{e.gdef1}
\end{equation}
The $g_{j/h}(x,\Tsc{b}{})$ function depends on the choice of input scale $Q_0$ and on the form of $\bstarsc$, but it is otherwise scale-independent. 

As long as there is negligible sensitivity to the very small $\Tsc{b}{} \ll 1/Q_0$ region, the form of \eref{expong} is sufficient to allow one to treat the overall factor $\tilde{f}_{j/h}(x,\bstarsc;\mu_{Q_0},Q_0^2)$ using collinear factorization. 
However, if we ultimately plan to evolve to such large $Q$ that the $\Tsc{b}{} \ll 1/Q_0$ limit becomes accessible, then we must transform the scales in $\tilde{f}_{j/h}(x,{\Tsc{b}{}}{};\mu_{Q_0},Q_0^2)$ once more by writing 
\begin{equation}
\tilde{f}_{j/h}(x,\bstarsc;\mu_{Q_0},Q_0^2) = \tilde{f}_{j/h}(x,\bstarsc;\mubstar,\mubstar^2) E(\mu_{Q_0}/\mubstar,Q_0/\mubstar,\mubstar \bstarsc) \, , \label{e.transs}
\end{equation}
where we have defined the scale 
\begin{equation}
\label{e.mubst}
\mu_{\bstarsc} \equiv C_1/\bstarsc \, ,
\end{equation}
and $C_1$ is a constant with size comparable to $1$. Then, $\tilde{f}_{j/h}(x,\bstarsc(\Tsc{b}{});\mubstar,\mubstar^2)$ is perturbatively very well behaved as $\Tsc{b}{} \to 0$.  
The factors in front of $e^{- g_{j/h}(x,\Tsc{b}{})}$ are never evaluated at $\Tsc{b}{}$ above $\bmax$, so one may choose $\bmax$ to be small enough that the $\order{\bmax \Lambda_{\rm QCD}}^a$ errors in \eref{ope} are negligible. Thus, we write \eref{expong} as
\begin{equation}
\tilde{f}_{{\rm OPE},j/h}(x,\bstarsc;\mubstar,\mubstar^2) E(\mu_{Q_0}/\mubstar,Q_0/\mubstar,\mubstar \bstarsc) e^{- g_{j/h}(x,\Tsc{b}{}) } + \order{\bmax \Lambda_{\rm QCD}}^a  \, , \label{e.transs2}
\end{equation}
where the ${\rm OPE}$ subscript symbolizes the first term in \eref{ope}. 

Isolating the nonperturbative part of the CS kernel is done in a way similar to the treatment of the TMD pdf itself. The evolution factor in the cross section, \eref{basicTMDmodelbspace2}, is rewritten in the form 
\begin{align}
& E(\mu_Q/\mu_{Q_0},Q/Q_0,\mu_{Q_0} \Tsc{b}{}) = 
E(\mu_Q/\mu_{Q_0},Q/Q_0,\mu_{Q_0} \bstarsc) \frac{E(\mu_Q/\mu_{Q_0},Q/Q_0,\mu_{Q_0} \Tsc{b}{})}{ E(\mu_Q/\mu_{Q_0},Q/Q_0,\mu_{Q_0} \bstarsc)}  \no
&{} \; = E(\mu_Q/\mu_{Q_0},Q/Q_0,\mu_{Q_0} \bstarsc) \exp \left\{ - \left[ \tilde{K}\parz{\bstarsc;\mu_{Q_0}} - \tilde{K}\parz{\Tsc{b}{};\mu_{Q_0}} \right] \ln \parz{\frac{Q}{Q_0}} \right\} \, . \label{e.Etrans}
\end{align}
Now one uses the fact (from \eref{TMDrg}) that scale dependence in \eref{Etrans} is $\Tsc{b}{}$-independent to identify and define the quantity in brackets as a scale-independent function called $g_K(\Tsc{b}{})$. So the evolution factor is
\begin{align}
E(\mu_Q/\mu_{Q_0},Q/Q_0,\mu_{Q_0} \Tsc{b}{}) &{}= 
E(\mu_Q/\mu_{Q_0},Q/Q_0,\mu_{Q_0} \bstarsc) 
\exp \left\{ - g_K(\Tsc{b}{}) \ln \parz{\frac{Q}{Q_0}} \right\} \, . \label{e.evolevol}
\end{align}
Now $E(\mu_Q/\mu_{Q_0},Q/Q_0,\mu_{Q_0} \bstarsc)$ is also never evaluated at $\Tsc{b}{} > \bmax$.\footnote{One more application of evolution equations is needed to convert the $\mu_{Q_0} \bstarsc$ to $\mubstar \bstarsc$ } 
Substituting \eref{transs2} and \eref{evolevol} into \eref{basicTMDmodelbspace2} reproduces the fully evolved cross section expression in the form that it is usually presented. 

In the above summary, all symbols are meant to be exact so, apart from possibly dropping the $\order{\bmax \Lambda_{\rm QCD}}^a$ in \eref{transs2}, nothing has been done to \eref{basicTMDmodelbspace2} other than to shuffle around factors arbitrarily. An implementation amounts to replacing these various pieces by approximations and parametrizations. The perturbatively calculable parts are normally obtained by truncating beyond some order, while in most approaches the nonperturbative functions $g_{j/h}(x,\Tsc{b}{})$  and $g_K(\Tsc{b}{})$ are replaced by nonperturbative ansatzes, which are then fit to data. 

Organizing the steps in this way has at least the following major advantages:
\begin{enumerate}[1.)]
\item The final $\tilde{f}_{j/h}(x,\bstarsc;\mu_{Q_0},Q_0^2)$ is restricted to perturbatively small values of $\Tsc{b}{}$. Calculating it using perturbative collinear factorization with the OPE involves no sensitivity to genuinely nonperturbative parameters, since
changing $\bstarsc(\Tsc{b}{})$ or the value of $\bmax$ only amounts to shifting contributions between $\tilde{f}_{j/h}(x,\bstarsc;\mu_{Q_0},Q_0^2)$ and $g_{j/h}(x,\Tsc{b}{})$. Analogous statements are true for the connection between $\tilde{K}(\bstarsc;\mu_{Q_0})$ and $g_K(\Tsc{b}{})$.

At least formally, therefore, the goal of isolating ``purely perturbative'' contributions is achieved with this prescription. In the final expression for the cross section, the only sensitivity to $\bstarsc(\Tsc{b}{})$ or $\bmax$ comes from dropping the $\order{\bmax \Lambda_{\rm QCD}}^a$ terms in \eref{transs2}, and these errors can be made arbitrarily small by choosing small enough $\bmax$. 

\item The $g$-functions, $g_{j/h}(x,\Tsc{b}{})$ and $g_K(\Tsc{b}{})$, that describe nonperturbative transverse coordinate dependence are exactly scale independent since they are defined with ratios and differences. The $g_{j/h}(x,\Tsc{b}{})$ function only depends on the initial choice of $Q_0$ (through the CS evolution), and both $g_{j/h}(x,\Tsc{b}{})$ and $g_K(\Tsc{b}{})$ depend on $\bstarsc$.  However, all $Q$-dependence is either in perturbatively calculable quantities or is just the $\ln Q$ multiplying $g_K(\Tsc{b}{})$.

\item There are formally no modifications to or approximations of the fundamental operator definitions.  For dealing with $\Tsc{b}{}$ becoming large, the only change is to the \emph{arguments} of $\tilde{f}_{j/h}(x,\Tsc{b}{};\mu_{Q_0},Q_0^2)$ and $\tilde{K}(\Tsc{b}{};\mu_{Q_0})$, not to the functions themselves. Thus, the nonperturbative functions $g_{j/h}(x,\Tsc{b}{})$ and $g_K(\Tsc{b}{})$ have clear and unambiguous relationships to their underlying operator definitions, namely \eref{gdef1} and $g_K(\Tsc{b}{}) = \tilde{K}\parz{\bstarsc;\mu_{Q_0}} - \tilde{K}\parz{\Tsc{b}{};\mu_{Q_0}}$.
\end{enumerate}

\section{Issues $\&$ complications}
\label{s.issues}

The advantages listed in 1.)-3.) above are formally true if the various parts of the TMD factorization expression for the cross section are understood to be exactly known. Of course, however, in practical implementations one always replaces them by approximations or ansatzes, and this is where complications may enter. Because $g_{j/h}(x,\Tsc{b}{})$ and $g_K(\Tsc{b}{})$ are where all the physics associated with nonperturbative transverse momentum dependence resides, and since detailed access to that information is the main goal in the present context, they will be the focus of the discussion below.
The main complications with the standard organizational scheme, from the perspective of hadron structure studies, are the following:
\begin{enumerate}[a)]
\item In general, if nonperturbative parametrizations for $g_{j/h}(x,\Tsc{b}{})$ and $g_K(\Tsc{b}{})$ are chosen arbitrarily, then item 1.) above will be at most only approximately true. If violations of 1.) are too large, then $\tilde{f}_{j/h}(x,\bstarsc;\mu_{Q_0},Q_0^2)$ is not actually sequestered from the nonperturbative parametrization. The $\bstarsc(\Tsc{b}{})$ function then effectively becomes an additional nonperturbative model and $\bmax$ becomes a model parameter. Therefore, nonperturbative model parameters impact the perturbative/collinear part, thus removing one of the motivating advantages of the $\bstarsc$ and $g$-function method.    

In applications to nonperturbative hadron structure, one of the tasks is to disentangle any purely nonperturbative transverse momentum dependence associated with bound states and hadronization from perturbative transverse momentum dependence, but violations of $\bstarsc$-independence indicate a breakdown in the ability to identify separate perturbative and nonperturbative parts.

\item In a parton-model-like picture, where TMD pdfs have a literal number density interpretation, the relationship between collinear and TMD pdfs is via simple integrals like 
\begin{equation}
f(x) = \int \diff[2]{\T{k}{}} f(x,\T{k}{}) \, .
\label{e.naive}
\end{equation}
In full QCD, this type of relation fails because the integral is ultraviolet (UV) divergent, but generalized versions of it~\cite{Rogers:2020tfs,Gonzalez-Hernandez:2022ifv,delRio:2024vvq} do hold in the form 
\begin{equation}
f(x;\mu_{Q}) = \int_{\rm reg.} \diff[2]{\T{k}{}} f(x,\T{k}{};\mu_{Q},\mu_Q^2) + \Delta + \text{p.s.} \, , \label{e.intrelext}
\end{equation}
where the ``${\rm reg}$'' subscript symbolizes a UV regulator that suppresses the contribution from $\Tsc{k}{} \gtrsim Q$, $\Delta$ is a higher order correction that is calculable in perturbative collinear factorization which is needed to transform the $\Tsc{k}{}$-regulated collinear pdf to the $\msbar$ scheme, and $\text{p.s.}$ means ``power-suppressed.'' An arbitrarily chosen functional form for $g_{j/h}(x,\Tsc{b}{})$ will not generally give a $\tilde{f}_{j/h}(x,{\Tsc{b}{}}{};\mu_{Q_0},Q_0^2)$ that satisfies \eref{intrelext} with previously known collinear pdf extractions for the left-hand side.

Equations of the form of \eqref{e.naive} are an important part of the hadron structure framework of interpretations, where the nonperturbative parts of TMD pdfs are understood to describe properties of bound state quark and gluon constituents. This is emphasized in the classic ``prism'' diagram that one frequently finds in literature on nonperturbative parton structure --- see \fref{prism}.
\begin{figure}[h!]
\centering
\includegraphics[width=8cm]
{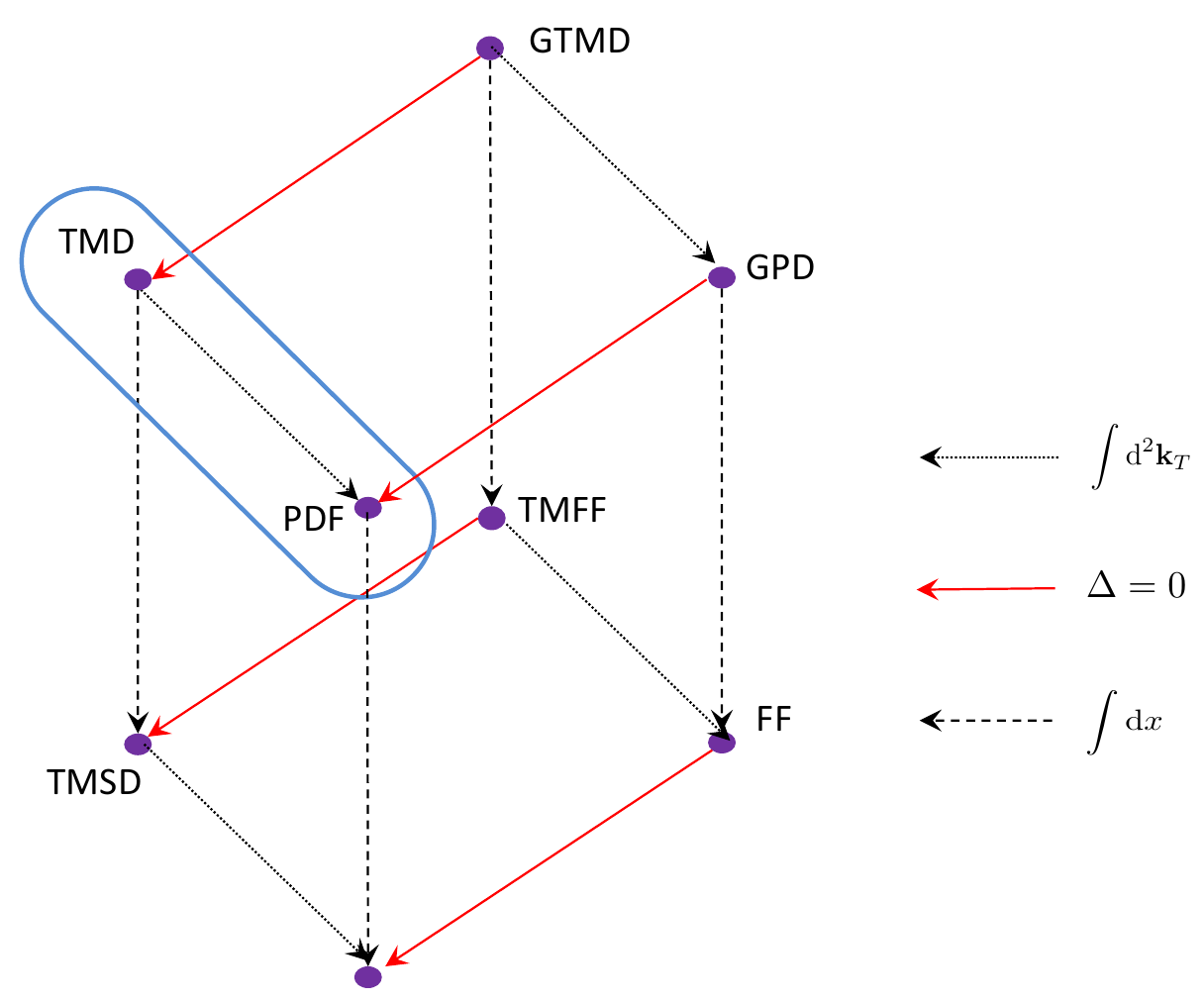}
\caption{A reproduction of the classic ``prism'' diagram from Ref.~\cite[figure 1]{Lorce:2011dv} showing the connections between different types of parton correlation functions in a nonperturbative structure interpretation. The transverse momentum integrals are typically UV divergent, and so preserving the interpretation in full QCD requires translating them into forms like \eref{intrelext}.}
\label{f.prism}
\end{figure}

Therefore, allowing the nonperturbative parts contained in $g_{j/h}(x,\Tsc{b}{})$ to be unconstrained by the generalized version of these integrals in \eref{intrelext} conflicts with a goal of isolating nonperturbative parts that are to be identified and interpreted with these kinds of nonperturbative structures in mind. In the standard way of implementing the $\bstarsc$/g-function prescription, there is not an obvious or direct way to ensure that $g_{j/h}(x,\Tsc{b}{})$ is subjected to such constraints.

\item In the usual way that the $\bstarsc$-method is implemented, the value of $\bmax$ actually plays two separate roles simultaneously: i.) It is treated as the $\Tsc{b}{}$ \emph{above} which transverse coordinate dependence is understood to become nonperturbative, and ii.) it is the $\Tsc{b}{}$ $\emph{below}$ which one transforms scales from $Q_0$ to $\mubstar \sim 1/\Tsc{b}{}$. But these two transition points have different physical meanings. Transition i.) deals with the physically meaningful question of where and how the completely nonperturbative transverse coordinate dependence starts to become important. It is part of the nonperturbative physics that one hopes to model or extract in phenomenological applications. Transition ii.), by contrast, is an arbitrary scheme change, and in principle it has no effect on exact quantities.  The details of how ii.) gets implemented are instead to be guided by the goal of optimizing convergence in perturbation theory calculations and minimizing higher orders. 

There is no reason in principle that these two transition points need to be exactly the same. However, the $\bstarsc$-method, at least in the specific way that it is usually set up, makes them identical. This puts unnecessarily rigid constraints on how one describes the nonperturbative-to-perturbative transition.

One may try to avoid this by introducing two separate $\bstarsc$ functions, with two separate $\bmax$'s. One of them can then be used to regulate the $\Tsc{b}{}$ argument at large $\Tsc{b}{}$ and the other can be used to construct $\mubstar$.  But that 
exacerbates the problem in a) by putting multiple arbitrary functions and parameters into the collinear factorization parts. The problem cannot be completely avoided because it is a consequence of relying on the $\Tsc{b}{}$-argument substitution method as a way to simultaneously regulate large $\Tsc{b}{}$ behavior where nonperturbative physics sets in and to transform to $\sim 1/\Tsc{b}{}$ scaling in the small $\Tsc{b}{} \ll 1/Q_0$ limit. 

\end{enumerate}

These complications are ultimately connected to the fact that the usual $\bstarsc$ organization of factors views $\Tsc{b}{}$ as coming from only two very sharply defined contributions, one that involves entirely perturbative $\Tsc{b}{}$-dependence and one that is entirely nonperturbative $\Tsc{b}{}$-dependence. However, a more realistic view is that there are three types of $\Tsc{b}{}$-dependence: 
\begin{enumerate}
\item Totally nonperturbative behavior as $\Tsc{b}{} \to \infty$.
\item Very reliable collinear factorization as $\Tsc{b}{} \ll 1/Q_0$.
\item An intermediate transition region around $\Tsc{b}{} \approx 1/Q_0$ where $\Tsc{b}{}$-dependence is reasonably well-described by perturbative collinear factorization, but is not as isolated from nonperturbative effects as region $\#2$.
\end{enumerate}

In an ideal TMD parametrization, region 3 above would be described by a physically motivated model that interpolates between the first two. However, the standard way of organizing the $\bstarsc$ prescription leads one to assign all $\Tsc{b}{}$-dependence only to one of either the first two categories.  The practical consequences of this are possibly not very serious if $Q$ is so large that the $g$-functions make a relatively small contribution that can be viewed as a correction to resummation calculations that only involves collinear pdfs. For this situation, it is possible that only the general qualitative features of the $g$-functions are important.

However, it becomes more of an issue when extracting the details of nonperturbative transverse momentum dependence and separating them from perturbatively calculable parts is the main target goal. Of course, one way around the problem is simply to construct parametrizations of the $g$-functions from the outset such that they explicitly evade a)-c) above, and this is essentially the approach we advocate below. (Although, an outcome is that most of the practical advantages of using a $\bstarsc$ prescription and $g$-function style of organization to start with are no longer present.)

The identification of the issues summarized above is not new. Our contribution is only to offer one strategy for systematizing their treatment. 
In particular, the problems that arise from nonperturbative model dependence affecting perturbative calculations have been discussed for a long time. 
References~\cite{Qiu:2000ga,Grewal:2020hoc}, for example, deal with the problem by imposing continuity and smoothness conditions at the boundary between the perturbative and nonperturbative regions, thus preserving more of the predictive power coming from the collinear factorization parts of calculations.

\section{Proposed modifications}
\label{s.mods}

In Refs.~\cite{Gonzalez-Hernandez:2022ifv,Gonzalez-Hernandez:2023iso,Aslan:2024nqg}, we proposed a set of steps designed to evade issues a)-c) above. Since they are meant for applications where extracting the details of nonperturbative transverse momentum (the behavior that is normally contained entirely in the $g$-functions) is the main objective, we have been calling it a ``hadron structure oriented'' (HSO) approach to TMD phenomenology. While Refs.~\cite{Gonzalez-Hernandez:2022ifv,Gonzalez-Hernandez:2023iso,Aslan:2024nqg} list many details, the overall basic strategy is simple to state:
\begin{enumerate}

\item Construct a smooth and continuous parametrization of $\tilde{f}_{j/h}(x,{\Tsc{b}{}}{};\mu_{Q_0},Q_0^2)$ that directly interpolates between perturbative and nonperturbative $\Tsc{b}{}$ descriptions for $\Tsc{b}{} \approx 1/Q_0$ and $\Tsc{b}{} \gg 1/Q_0$ respectively.\footnote{Refs.~\cite{Gonzalez-Hernandez:2022ifv,Gonzalez-Hernandez:2023iso,Aslan:2024nqg} actually interpolated in $\Tsc{k}{}$-space, but the basic strategy is the same in both transverse coordinate or momentum space.} 

\item Impose the constraint in \eref{intrelext} on this parametrization. 

\item Ensure that the parametrization includes the usual scale transformation for the $\Tsc{b}{} \ll 1/Q_0$ region that transforms $\mu_{Q_0}$,$Q_0$ to $\sim 1/\Tsc{b}{}$. 

\item Perform analogous steps for $\tilde{K}(\Tsc{b}{};\mu_{Q_0})$.

\item Evolve to $Q > Q_0$ for comparing with higher scales.

\end{enumerate}

Once this is done, the result can be transformed into the usual $\bstarsc$/$g$-function construction if that is desired. The resulting $g$-functions will no longer have the complications listed in a)-c) of the previous section.

\section{Questions}

The following questions arose in the context of the talk:

\subsection{Can one use a $\bmin$ regulator rather than a transverse momentum cutoff?}
\label{s.example}

References~\cite{Gonzalez-Hernandez:2022ifv,Gonzalez-Hernandez:2023iso,Aslan:2024nqg} used a direct transverse momentum cutoff to regulate the integral in \eref{intrelext}, with an input parametrization constructed in transverse momentum space. This was for various practical reasons, but it is not essential to the basic procedure, and other regulators are acceptable. 
Another common approach, especially for parametrizations constructed directly in $\Tsc{b}{}$-space, is to freeze $\Tsc{b}{}$ below some $\bmin$ to regulate the integral.
It is instructive to repeat steps analogous to Secs.~III and IV of Refs.~\cite{Aslan:2024nqg}, but now in $\Tsc{b}{}$-space directly and with a small $\Tsc{b}{}$ cutoff to regulate the UV divergence rather than a large $\Tsc{k}{}$ cutoff in transverse momentum space. 

First, one defines the cutoff regulator prescription by specifying an arbitrary cutoff function, 
\begin{equation}
b_{c}(\Tsc{b}{}) = 
\begin{cases}
\Tsc{b}{} & \Tsc{b}{} \gtrsim \bmin \\
\bmin & \Tsc{b}{} \ll \bmin
\end{cases} \, , 
\end{equation}
with $\bmin \approx 1/\mu_{Q_0}$. It could be any function with this property, but for concreteness let us choose
\begin{equation}
b_c(\Tsc{b}{}) = \sqrt{\Tscsq{b}{} + \bmin^2} \, .
\end{equation}
Then the integral in \eref{intrelext} is regulated if we replace the $\Tsc{b}{}$-space TMD pdf by one with the $\Tsc{b}{}$ argument replaced by $b_{c}(\Tsc{b}{})$,
\begin{equation}
\label{e.replace}
\tilde{f}_{j/h}(x,{\Tsc{b}{}}{};\mu_{Q_0},Q_0^2) \longrightarrow
 \tilde{f}_{j/h}(x,b_{c}(\Tsc{b}{});\mu_{Q_0},Q_0^2) \, . 
\end{equation}

In $\Tsc{b}{}$-space, step 1 of~\sref{mods} would be to construct a parametrization that interpolates between the logarithmically diverging perturbative collinear factorization prediction at small $\Tsc{b}{}$ and the nonperturbative behavior at large $\Tsc{b}{} \to \infty$. If we keep only the lowest nontrivial order at small $\Tsc{b}{}$, this would mean we need
\begin{align}
&{} \tilde{f}(x,\Tsc{b}{};\mu_{Q_0};Q_0^2)_\text{param} \no
&{}= 
\begin{cases}
\text{Constant} - A \ln\parz{\frac{\Tsc{b}{} \mu_{Q_0}}{b_0}} - B \left[ \ln^2\parz{\frac{\Tsc{b}{} \mu_{Q_0}}{b_0}} + \ln\parz{\frac{\Tsc{b}{} \mu_{Q_0}}{b_0}} \ln \parz{\frac{Q_0^2
}{\mu_{Q_0}^2}} \right]& \Tsc{b}{} \ll 1/\Lambda_{\rm QCD} \\
\text{nonperturbative parametrization} & \Tsc{b}{} \to \infty
\end{cases} \, ,
\end{align}
where $b_0 = 2 e^{-\gamma_E}$, $A$ and $B$ are known $\order{\alpha_s}$ functions of $x$ calculated in collinear perturbation theory and the constant term is 
ultimately going to be fixed by imposing \eref{intrelext}, which for  the $\bmin$ cutoff is
\begin{equation}
\label{e.intrelb}
\int \diff[2]{\T{k}{}} \int \frac{\diff[2]{\T{b}{}}}{(2 \pi)^2} e^{i \T{k}{} \cdot \T{b}{}} \tilde{f}(x,b_c(\Tsc{b}{});\mu_{Q_0};Q_0^2)_\text{param} = \tilde{f}(x,\bmin;\mu_{Q_0};Q_0^2)_\text{param} = f(x;\mu_{Q_0}) - \Delta \, .
\end{equation}
An example of a reasonable parametrization that satisfies the above might be 
\begin{align}
&{}\tilde{f}(x,\Tsc{b}{};\mu_{Q_0};Q_0^2)_\text{param} \no
&{} \; = C e^{-m_a \Tscsq{b}{}}  - A \ln\parz{\frac{\Tsc{b}{} \mu_{Q_0}}{b_0 + m_b \Tsc{b}{} }} - B \left[ \ln^2\parz{\frac{\Tsc{b}{} \mu_{Q_0}}{b_0 + m_b \Tsc{b}{}}} + \ln\parz{\frac{\Tsc{b}{} \mu_{Q_0}}{b_0 + m_b \Tsc{b}{}}} \ln \parz{\frac{Q_0^2
}{\mu_{Q_0}^2}} \right]  \no
&{} \; + A \ln\parz{\frac{\mu_{Q_0}}{m_b}} + B \left[ \ln^2\parz{\frac{\mu_{Q_0}}{m_b}} + \ln\parz{\frac{\mu_{Q_0}}{m_b}} \ln \parz{\frac{Q_0^2
}{\mu_{Q_0}^2}} \right] \, . \label{e.bparam}
\end{align}
The nonperturbative model parameters are the mass scales $m_a$ and $m_b$, which are expected to be comparable in size to $\Lambda_{\rm QCD}$. The model vanishes as $\Tsc{b}{} \to \infty$, so the Fourier-Bessel transform to transverse momentum space exists as an ordinary function. $C$ is the constant term that will be fixed by imposing \eref{intrelb}:
\begin{align}
&{} C e^{-m_a \bmin^2}  - A \ln\parz{\frac{\bmin \mu_{Q_0}}{b_0 + m_b \bmin }} - B \left[ \ln^2\parz{\frac{\bmin \mu_{Q_0}}{b_0 + m_b \bmin}} + \ln\parz{\frac{\bmin \mu_{Q_0}}{b_0 + m_b \bmin}} \ln \parz{\frac{Q_0^2
}{\mu_{Q_0}^2}} \right]  \no
&{} + A \ln\parz{\frac{\mu_{Q_0}}{m_b}} + B \left[ \ln^2\parz{\frac{\mu_{Q_0}}{m_b}} + \ln\parz{\frac{\mu_{Q_0}}{m_b}} \ln \parz{\frac{Q_0^2
}{\mu_{Q_0}^2}} \right] = f(x;\mu_{Q_0}) - \Delta \, ,
\label{e.intrelb2}
\end{align}
which fixes 
\begin{align}
 C =&{} f(x;\mu_{Q_0}) - \Delta + A \ln\parz{\frac{\bmin \mu_{Q_0}}{b_0}} + B \left[ \ln^2\parz{\frac{\bmin \mu_{Q_0}}{b_0}} + \ln\parz{\frac{\bmin \mu_{Q_0}}{b_0}} \ln \parz{\frac{Q_0^2
}{\mu_{Q_0}^2}} \right]  \no
&{} - A \ln\parz{\frac{\mu_{Q_0}}{m_b}} - B \left[ \ln^2\parz{\frac{\mu_{Q_0}}{m_b}} + \ln\parz{\frac{\mu_{Q_0}}{m_b}} \ln \parz{\frac{Q_0^2
}{\mu_{Q_0}^2}} \right] + \text{p.s.} \, ,
\label{e.intrelb3}
\end{align}
where the ``$\text{p.s.}$'' terms are $\order{\bmin \Lambda_{\rm QCD}}$. Fixing $C$ to the leading power terms in \eref{intrelb3} amounts to step 2 of \sref{mods}. 
Recall that we are considering the limit that $\Lambda_{\rm QCD}/\mu_{Q_0}$ is small, so the terms on the second line in \eref{intrelb3} are not perturbatively suppressed even though the $A$ and $B$ factors  are proportional to $\alpha_s(\mu_{Q_0})$. Setting $\bmin = b_0/\mu_{Q_0}$ eliminates the logarithmic terms on the first line. Then, the full parametrization in \eref{bparam} is 
\begin{align}
&{} \tilde{f}(x,\Tsc{b}{};\mu_{Q_0};Q_0^2)_\text{param} \no
&{} \; = e^{-m_a \Tscsq{b}{}} \left\{ f(x;\mu_{Q_0}) - \Delta  - A \ln\parz{\frac{\mu_{Q_0}}{m_b}} - B \left[ \ln^2\parz{\frac{\mu_{Q_0}}{m_b}} + \ln\parz{\frac{\mu_{Q_0}}{m_b}} \ln \parz{\frac{Q_0^2
}{\mu_{Q_0}^2}} \right] \right\}   \no 
&{} \;  - A \ln\parz{\frac{\Tsc{b}{} \mu_{Q_0}}{b_0 + m_b \Tsc{b}{} }} - B \left[ \ln^2\parz{\frac{\Tsc{b}{} \mu_{Q_0}}{b_0 + m_b \Tsc{b}{}}} + \ln\parz{\frac{\Tsc{b}{} \mu_{Q_0}}{b_0 + m_b \Tsc{b}{}}} \ln \parz{\frac{Q_0^2
}{\mu_{Q_0}^2}} \right]  \no
&{} \; + A \ln\parz{\frac{\mu_{Q_0}}{m_b}} + B \left[ \ln^2\parz{\frac{\mu_{Q_0}}{m_b}} + \ln\parz{\frac{\mu_{Q_0}}{m_b}} \ln \parz{\frac{Q_0^2
}{\mu_{Q_0}^2}} \right] \, . \label{e.bparamfinal}
\end{align}
Now the limit of $\Tsc{b}{} \Lambda_{\rm QCD} \to 0$ reproduces the usual well-known OPE expression,
\begin{multline}
\tilde{f}(x,\Tsc{b}{};\mu_{Q_0};Q_0^2)_\text{param} \stackrel{\Tsc{b}{} \Lambda_{\rm QCD} \to 0}{=} \\ 
f(x;\mu_{Q_0}) - \Delta  - A \ln\parz{\frac{\Tsc{b}{} \mu_{Q_0}}{b_0}} - B \left[ \ln^2\parz{\frac{\Tsc{b}{} \mu_{Q_0}}{b_0}} + \ln\parz{\frac{\Tsc{b}{} \mu_{Q_0}}{b_0}} \ln \parz{\frac{Q_0^2
}{\mu_{Q_0}^2}} \right] + \order{\Tsc{b}{} \Lambda_{\rm QCD}}^a \, ,
\end{multline}
and \eref{intrelext} holds by construction with only $\order{\bmin \Lambda_{\rm QCD}}^a$ errors. The large $\Tsc{b}{}$ behavior in this model is essentially Gaussian. The constant terms in the OPE arise from the $\Delta$ term. 

For dealing with the cross section behavior at $Q \approx Q_0$ through $\order{\alpha_s(\mu_{Q_0})}$, this is sufficient. However, if we plan to evolve to $Q \gg Q_0$, then the $\Tsc{b}{} \ll 1/Q_0$ region also needs to be properly accounted for. To deal with this, it is straightforward to perform another scale transformation from $\mu_{Q_0}, Q_0$ to $\sim 1/\Tsc{b}{}$ using the evolution equations in \erefs{CSeq}{TMDrg} again. Since this transformation takes place in the $\Tsc{b}{} \lesssim 1/Q_0$ region of coordinate space, it is under control perturbatively and the errors it introduces can be minimized with detailed treatments of higher orders in perturbation theory. 

If it is preferred, one may now take this parametrization and 
construct a $g_{j/h}(x,\Tsc{b}{})$ function just by substituting it into \erefs{firstratio}{gdef1},
\begin{equation}
\tilde{f}(x,\Tsc{b}{};\mu_{Q_0};Q_0^2)_\text{param} = \tilde{f}_{j/h}(x,\bstarsc;\mu_{Q_0},Q_0^2)_\text{param} e^{- g_{j/h}(x,\Tsc{b}{})_\text{param} }  \,  \label{e.expong2}
\end{equation}
with
\begin{equation}
g_{j/h}(x,\Tsc{b}{})_\text{param} \equiv - \ln \parz{\frac{\tilde{f}_{j/h}(x,{\Tsc{b}{}}{};\mu_{Q_0},Q_0^2)_\text{param}}{\tilde{f}_{j/h}(x,\bstarsc;\mu_{Q_0},Q_0^2)_\text{param}}} \, . \label{e.gdef12}
\end{equation}
However, now this $g_{j/h}(x,\Tsc{b}{})_\text{param}$ evades all the complications listed in a)-c) of \sref{issues}. In particular, there is no disadvantage in making $\bmax$ conservatively very small, $\bmax \lesssim 1/Q_0$. The consequence of doing so is only that a significant part of the perturbative transverse coordinate dependence is handled by 
$g_{j/h}(x,\Tsc{b}{})_\text{param}$. Therefore, one may ensure that the power suppressed $\order{\Lambda_{\rm QCD} \bmax}^a$ terms
in \eref{transs2} are negligible without sacrificing the predictive power coming from collinear factorization. The $\bstarsc$-independence in item 1 of \sref{standard} is exactly realized in a $g_{j/h}(x,\Tsc{b}{})_\text{param}$ constructed through steps like the above. 

It is instructive to contrast this with the kind of result that would be obtained if one followed more typical steps.  Let us label such a result as ``$\text{param,2}$.'' There, one writes \eref{expong} with $\tilde{f}_{j/h}(x,\bstarsc;\mu_{Q_0},Q_0^2)$ immediately replaced by its OPE expression, 
\begin{equation}
\tilde{f}_{j/h}(x,{\Tsc{b}{}}{};\mu_{Q_0},Q_0^2)_{\text{param,2}} = \tilde{f}_{{\rm OPE},j/h}(x,\bstarsc;\mu_{Q_0},Q_0^2) e^{- g_{j/h}(x,\Tsc{b}{})_{\text{param,2}} }  \, , \label{e.expong3}
\end{equation}
and $g_{j/h}(x,\Tsc{b}{})_{\text{param,2}}$ is replaced with a simple ansatz. A typical example might be a quadratic power law,
\begin{equation}
-g_{j/h}(x,\Tsc{b}{})_{\text{param,2}} = - m_a^2 \Tscsq{b}{} \, .
\end{equation}
Now the $\bstarsc$-independence is clearly not preserved to very high accuracy because this would require at least that 
\begin{equation}
e^{-g_{j/h}(x,\Tsc{b}{})_{\text{param,2}}} \approx 
- \ln \parz{\frac{\tilde{f}_{{\rm OPE},j/h}(x,{\Tsc{b}{}}{};\mu_{Q_0},Q_0^2)}{\tilde{f}_{{\rm OPE},j/h}(x,\bstarsc;\mu_{Q_0},Q_0^2)}} \label{e.param2g}
\end{equation}
in the region of small $\Tsc{b}{} \approx \bmax$. For a power law $g_{j/h}(x,\Tsc{b}{})$, however, such a correspondence can only be a rough approximation within some narrow range of moderate $\Tsc{b}{}$. 

Furthermore, 
applying \eref{intrelb} to \eref{expong3} gives 
\begin{align}
&{}\int \diff[2]{\T{k}{}} \int \frac{\diff[2]{\T{b}{}}}{(2 \pi)^2} e^{i \T{k}{} \cdot \T{b}{}} \tilde{f}\parz{x,b_c(\Tsc{b}{});\mu_{Q_0};Q_0^2}_{\text{param,2}}\no
&{} \qquad = \tilde{f}_{{\rm OPE},j/h}(x,\bstarsc(\Tsc{b}{} = \bmin);\mu_{Q_0},Q_0^2) e^{- m_a^2 \bmin^2 } \no
&{} \qquad = \tilde{f}_{{\rm OPE},j/h}(x,\bmin;\mu_{Q_0},Q_0^2) + \order{\frac{\bmin^2}{\bmax^2}} + \order{\Lambda_{\rm QCD}^2 \bmin^2} \, \no
&{} \qquad = f(x;\mu_{Q_0}) - \Delta + \order{\frac{\bmin^2}{\bmax^2}} + \order{\Lambda_{\rm QCD}^2 \bmin^2} \, .
\end{align}
However, near the input scale both $\bmin$ and $\bmax$ are $\order{1/Q_0}$, so the $\order{\bmin^2/\bmax^2}$ term is \emph{not} a power suppressed correction (and so the expansion in $\bmin/\bmax$ is not valid). Therefore, \eref{intrelb} fails in general for \eref{expong3}. To be preserved, the $g_{j/h}(x,\Tsc{b}{})$ function needs to be constructed so as to cancel the  $\order{\bmin^2/\bmax^2}$ behavior. 

After evolution to $Q \gg Q_0$, the $\bmin$ becomes $\sim 1/Q$ while $\bmax$ remains $\sim 1/Q_0$. Then, the $\order{\bmin^2/\bmax^2}$ error terms are $\order{Q_0^2/Q^2}$, which are eventually negligible as $Q$ increases. However, much of the data of interest for studies of hadron structure are in the region of $Q \approx Q_0$. 

This example of a model parametrization, \eref{bparamfinal}, is only meant for illustration purposes and others are probably much better in practice. The point of this discussion is not the details of the parametrization itself, but rather the general steps for constructing it.  

\subsection{Is \eref{intrelext} automatically satisfied within the standard organization?}
\label{s.intrelquestion}

Equation~\eqref{e.intrelext} is obviously not guaranteed to be satisfied for arbitrary functions substituted into the integrand, given a fixed extraction for an $\msbar$ collinear pdf on the left-hand side. It will be satisfied eventually for large enough $Q$ for any parametrization that matches to the OPE as $\Tsc{b}{} \to 0$.
That includes standard existing parametrizations that use the usual $\bstarsc$-procedure as described in \sref{standard}. However, the reason is that at large enough $Q$ the sensitivity of the $\Tsc{k}{}$-integral to the $g$-functions eventually vanishes. Specifically, the version of \eref{intrelext} that arises automatically in a standard $\bstarsc$ implementation has the form, 
\begin{equation}
f(x;\mu_{Q}) = \int_{\rm reg.} \diff[2]{\T{k}{}} f(x,\T{k}{};\mu_{Q},\mu_Q^2) + \Delta + \order{\frac{Q_0}{Q}}^a \, , \qquad a > 0 \, .\label{e.intrelext2}
\end{equation}
In other words, the power suppressed error terms are in the form of ratios of the input scale to the hard scale, so the relation fails for applications at $Q \approx Q_0$, which is where sensitivity to nonperturbative parts is the greatest. As the illustrative example above shows, this is the result if one does not incorporate the $g$-functions along with the perturbative parts in the integral condition.
\begin{figure}[h!]
\centering
\includegraphics[width=10cm]
{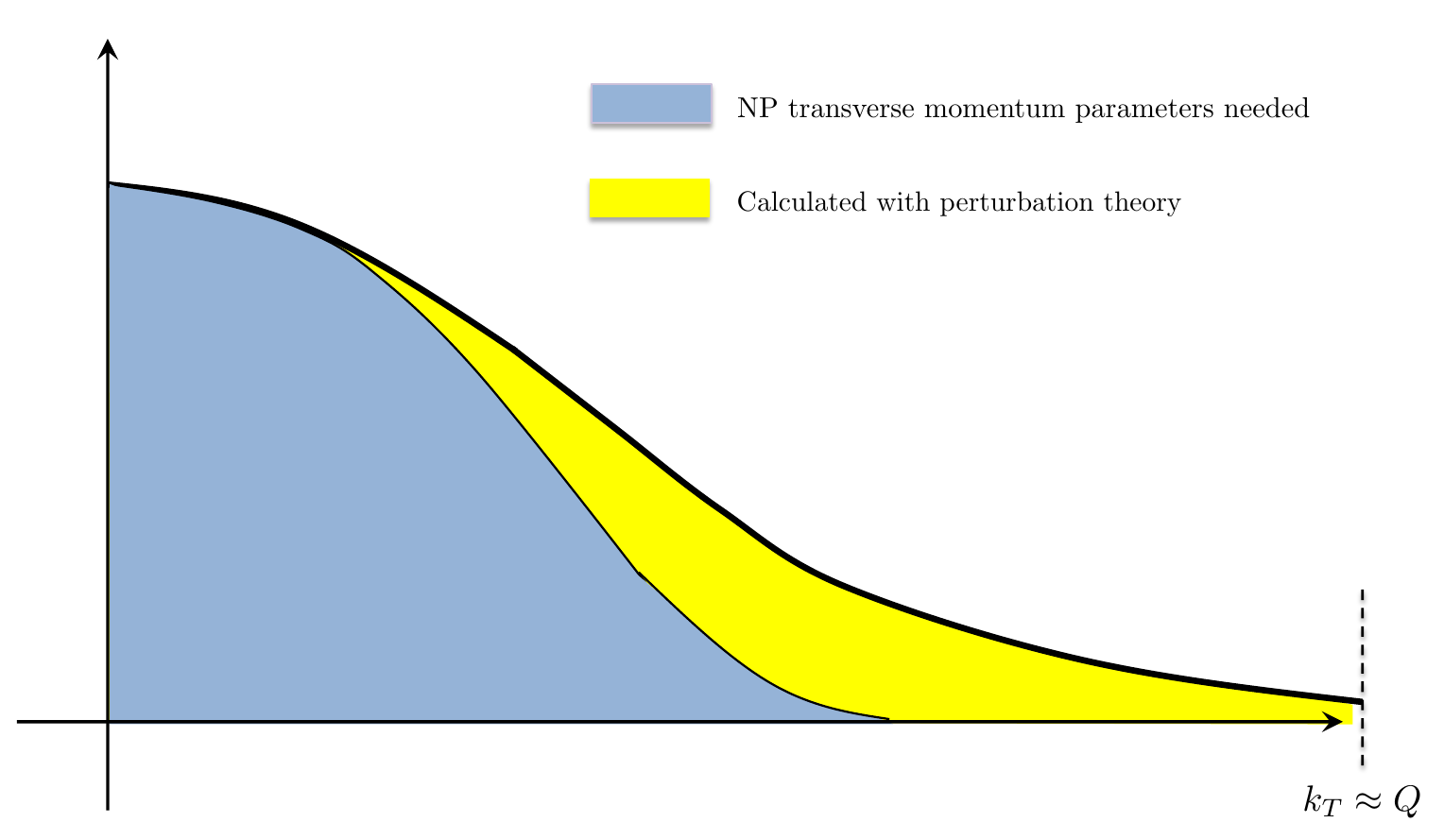}
\caption{Schematic representation of the relative contributions from perturbative (yellow) and nonperturbative (blue) transverse momentum dependence to the $\Tsc{k}{}$-integral of a TMD pdf. For large enough $Q$, the yellow contribution dominates.}
\label{f.pvnp}
\end{figure}

Intuitively, the reason it eventually holds for $Q \gg Q_0$ is that the integrand in \eref{intrelext2} becomes completely dominated by a perturbative tail whose description only involves collinear pdfs and none of the sensitivity to nonperturbative transverse momentum dependence. Figure~\ref{f.pvnp} provides a schematic visualization of this. The curve is meant to represent a TMD pdf plotted versus $\Tsc{k}{}$, and the shaded blue and yellow areas are meant to represent the contributions from nonperturbative and perturbative transverse momentum dependence respectively to the integral over $\T{k}{}$, assuming a regulator that removes the $\Tsc{k}{} \gtrsim Q$ region. As $Q$ grows, the contribution from the yellow part increases until there is essentially no non-negligible contribution from the blue part. That is, the area under the curve is completely dominated by behavior that is describable from the outside in terms of collinear pdfs. 

But situations where nonperturbative transverse momentum plays no role are the opposite of what one is seeking in the context of nonperturbative hadron structure applications. While it might turn out that for specific large values of $Q$ it is the case that $Q_0/Q$ is small enough that the error terms in \eref{intrelext2} are negligible, assuming this is true from the outset restricts one to considering only conservatively very large $Q$, and it contradicts the goal of extracting the nonperturbative transverse momentum dependence associated with bound states or hadronization. Furthermore, it discards a valuable consistency constraint on the $g$-functions, which could be important for the $\Tsc{k}{} \ll Q$ region of the differential cross section even at larger $Q$.
What is needed for applications where the details of the $g$-functions are important is a version of \eref{intrelext2} of the form
\begin{equation}
f(x;\mu_{Q}) = \int_{\rm reg.} \diff[2]{\T{k}{}} f(x,\T{k}{};\mu_{Q},\mu_Q^2) + \Delta + \order{\frac{\Lambda_{\rm QCD}}{Q_0}}^a \, , \qquad a > 0 \, \label{e.intrelext3}
\end{equation}
so that the integrals relating TMD and collinear pdfs hold even for $Q \approx Q_0$. This is not automatically satisfied for arbitrary parametrizations of $g_{j/h}(x,\Tsc{b}{})$ but rather needs to be imposed on them directly. For example, it is satisfied by construction in the example in \eref{bparamfinal} but not in \eref{expong3}.

Examples in toy model field theories are also instructive for illustrating the effect. In Ref.~\cite{Aslan:2022zkz}, for example, illustrative examples with a scalar Yukawa theory found that significant violations of \eref{intrelext3} with a cutoff regulator result if the standard $g$-function ansatz method is used.

\subsection{What is the difference with the $\mubstar$ approach?}

Both the more traditional organization of TMD factorization and the approach we have adopted in \sref{mods} involve a scheme transformation to $\sim 1/\Tsc{b}{}$ behavior at $\Tsc{b}{} \ll 1/Q_0$. See, for example, the function named $\overline{Q}_0(\Tsc{b}{},a)$ in Eq.(77) of Ref.~\cite{Aslan:2024nqg}, which is similar to \eref{mubst}. This can give the appearance that there is no real difference between the two methods. 

The significant difference is related to issue c) of \sref{issues}. Recall that in the example of \sref{example} the nonperturbative transverse momentum dependence is controlled entirely by the model parameters $m_a$ and $m_b$. 
Changing the scales to $\sim 1/\Tsc{b}{}$ is left as a separate step, and can be chosen to occur at much large values of $1/\Tsc{b}{}$ than $m_a$ or $m_b$. The form of the function used in the scale transformation, and the value of $\Tsc{b}{}$ where it is implemented, do not need to be connected in any way with how one models the nonperturbative transverse momentum dependence as $\Tsc{b}{} \to \infty$. This separation is the main difference from the usual approach. 

Contrast this with a parametrization like \erefs{expong3}{param2g}. If, in attempting to make $\order{\alpha_s(\mubstar)^n}$ and $\order{\Lambda_{\rm QCD} \bmax}^a$ errors as small as possible, one makes $\bmax$ too small, then one will be forced to describe even the perturbatively describable $\Tsc{b}{}$-dependence with a nonperturbative ansatz. If, in attempting to widen the range that is described perturbatively as far as possible, one makes $\bmax$ too large, then $\alpha_s(\mubstar)$ becomes a nonperturbative large coupling and $\Lambda_{\rm QCD} \bmax$ errors stop being negligible. Notice the practical difficulties that this trade-off creates for disentangling truly nonperturbative ($\Tsc{b}{} > 1/Q_0$) from perturbative ($\Tsc{b}{} \ll 1/Q_0$) transverse momentum dependence. 

We argue that such a trade-off is not actually necessary and is only an artifact of using an overly restrictive organizational scheme.  

\subsection{Is it useful to keep the TMD pdfs evaluated at a $b_c(\Tsc{b}{})$?}

The TMD pdfs in the $\msbar$ scheme and with fixed $\mu = \sqrt{\zeta} \sim Q_0$ are determined by their operator definitions for all $\Tsc{b}{}$, and therefore, from the evolution equations, for all higher $Q$. In particular, assuming that the $\msbar$ pdfs are known, the small $\Tsc{b}{}$ (or large-$\Tsc{k}{}$) parts are uniquely determined by the OPE, and this rigidity is an important part of the predictive power of the overall formalism.  If a replacement like \eref{replace} is made and kept, then the TMD pdfs no longer follow the OPE and they no longer match their operator definition at a given value of $\Tsc{b}{}$. This can be seen directly in the example above. In \eref{bparamfinal}, if the replacement $\Tsc{b}{} \to b_c(\Tsc{b}{}) = \sqrt{\Tscsq{b}{} + \bmin^2}$ is made, then a TMD pdf does not transform into its OPE expression as $\Tsc{b}{} \to 0$, but rather freezes at its value around $\Tsc{b}{} \approx \bmin$. Another way to see it directly is by considering the Fourier-Bessel transforms of common functional forms. The logarithms of the OPE in $\Tsc{b}{}$-space translate into $(1/\Tscsq{k}{}) \times \text{logarithms}$ in $\Tsc{k}{}$-space:
\begin{equation}
    \ln\parz{\frac{\Tsc{b}{}Q}{2 e^{-\gamma_E}}}\mapsto -\frac{1}{2\pi \Tscsq{k}{}},\qquad  \ln^2\parz{\frac{\Tsc{b}{}Q}{2 e^{-\gamma_E}}}\mapsto -\frac{1}{2\pi \Tscsq{k}{}} \ln\parz{\frac{Q^2}{\Tscsq{k}{}}} \label{e.transformed}
\end{equation}
whereas with the $b_c(\Tsc{b}{})$ in place
\begin{align}
&\ln \parz{\frac{Q}{2 e^{-\gamma_E}} \sqrt{\Tscsq{b}{} + \bmin^2}} \mapsto -\frac{\bmin}{2\pi \Tsc{k}{}} K_1(\bmin \Tsc{k}{}) \, , \no 
& \ln^2 \parz{\frac{Q}{2 e^{-\gamma_E}} \sqrt{\Tscsq{b}{} + \bmin^2}} \mapsto -\frac{1}{2\pi \Tscsq{k}{}} \left[K_0(\bmin \Tsc{k}{}) + \Tsc{k}{} \bmin \ln \parz{\frac{\bmin Q^2}{2 e^{-\gamma_E}\Tsc{k}{}}} K_1(\bmin \Tsc{k}{}) \right] \, , 
\end{align}
which only matches \eref{transformed} in the limit that $\Tsc{k}{} \ll 1/\bmin$. For larger $Q$, one would typically choose a $\bmin \sim 1/Q$. Then, it could be argued that the only physically relevant region of $\Tsc{k}{}$ is anyway $\Tsc{k}{} \ll 1/\bmin$, so the $\Tsc{b}{} \to b_c(\Tsc{b}{})$ replacement should be irrelevant. However, the replacement nevertheless allows another adjustable arbitrary parameter to influence the small-to-large transverse momentum transition, so our view is that it should be avoided. 

In earlier work~\cite{Collins:2016hqq}, I advocated a procedure that did involve keeping the $\Tsc{b}{} \to b_c(\Tsc{b}{})$ replacement as described above in the hopes that having the freedom to shift contributions between the $W$-term and the $Y$-term would lead to a more refined treatment of large transverse momentum dependence in the cross section. However, because of the increase in adjustable parameters in the large transverse momentum region and the associated loss of predictive power, I no longer view this as an optimal approach. 

\section*{Acknowledgments}
T.~Rogers and T.~Rainaldi were supported by the U.S. Department of Energy, Office of Science, Office of Nuclear Physics, under Award Number DE-SC0024715. F.~Aslan was suppoprted by the Center for Nuclear Femtography, Southeastern Universities Research
Association, Washington, D.C..  T.~Rainaldi, T.~Rogers, and A.~Simonelli 
were supported by the DOE Contract No. DE- AC05-06OR23177, under which 
Jefferson Science Associates, LLC operates Jefferson Lab.
M. Boglione and J.O. Gonzalez-Hernandez acknowledge funding from the European
Union’s Horizon 2020 research and innovation programme under grant agreement No 824093 (STRONG2020), and from the European Union “Next
Generation EU” program through the Italian PRIN 2022 grant n. 20225ZHA7W.


\bibliographystyle{plainurl}
\bibliography{bibliography}

\end{document}